\documentclass{elsart}
\usepackage{amssymb}
\usepackage[]{fontenc}
\usepackage[latin1]{inputenc}
\usepackage{floatflt}
\usepackage{graphicx}
\begin{document}

\begin{frontmatter}

\title{MAGIC and the Search for Signatures of Supersymmetric Dark Matter}

\author{Dominik Elsässer\corauthref{cor1}} and   
\ead{elsaesser@astro.uni-wuerzburg.de}
\corauth[cor1]{Corresponding author.}
\author{Karl Mannheim}
\ead{mannheim@astro.uni-wuerzburg.de}
\address{Institut für Theoretische Physik und Astrophysik der Universität Würzburg}
\author{for the MAGIC Collaboration\thanksref{list}}
\thanks[list]{Full collaborators list can be found at the\\ collaboration website http://magic.mppmu.mpg.de}
\begin{abstract}
The 17m Imaging Air shower Cherenkov Telescope MAGIC (Roque de los Muchachos Observatory, La Palma, Canary Islands) has recently entered its commissioning phase. One of the main goals of the MAGIC telescope project is to provide an unprecedented sensitivity for the detection of gamma rays with energies as low as 30 GeV. A dedicated search for the gamma rays expected to be produced by WIMP annihilations is a prime object of the MAGIC physics program. We consider annihilating supersymmetric dark matter in M 87 and discuss a possible observation strategy. New calculations concerning the extragalactic gamma ray and neutrino backgrounds owing to cosmological neutralino annihilation are also briefly discussed.
\end{abstract}

\end{frontmatter}

\section{The MAGIC Telescope Project}
\label{}
The Major Atmospheric Gamma ray Imaging Cherenkov telescope is located on the Roque de los Muchachos, a site known for its excellent atmospheric quality for astronomy. Owing to the 17m diameter of the primary mirror and the high efficiency 577-pixel photomultiplier camera, MAGIC will provide the highest sensitivity below 300 GeV among the current generation of Imaging Air shower Cherenkov Telescopes (IACTs). Due to the larger collection area, MAGIC will also have a superior sensitivity above the energy threshold when compared with the planned GLAST satellite-experiment. Fast readout electronics and advanced analysis techniques have made ground-based observations below 100 GeV feasible. The effective area for air shower observations lies between $10^{4}$ and $10^{5}$ square meters. IACTs typically have a rather limited energy resolution, depending on zenith angle and energy, and for MAGIC varying between 15\% and 40\% as determined from Monte Carlo studies \cite{magic_overview,bretz}. Following the inauguration on October 10th, 2003, MAGIC has entered its commissioning phase. During commissioning, the first targets observed, and detected with high significance, were the known gamma ray sources Crab Nebula (M 1) and Markarian 421 \cite{first_alphas}.
Plans for the future are the construction of a second 17m telescope, which will commence in 2004, allowing for dedicated long-term observations, and the proposed European Cherenkov Observatory (ECO-1000) \cite{eco}, which would use a 34m central telescope to allow for observations down to about 10 GeV.

General physics goals for MAGIC include the full scope of high energy gamma ray astrophysics. Among the proposed targets for observations are pulsars and supernova remnants, Active Galactic Nuclei (AGN) and gamma ray bursts. Also, tests on quantum gravity effects and the cosmological gamma ray horizon due to pair creation in the metagalactic radiation field will be performed. Most notably, the low energy threshold shall open the view to more distant - and thus presumably more numerous - Active Galactic Nuclei. A MAGIC Dark Matter Working Group has been established \cite{eth} to optimize the search strategy for gamma rays due to the possible annihilation of Weakly Interacting Massive Particles (WIMPs).

\section{Gamma Rays from Neutralino Annihilations}
Supersymmetric extensions of the Standard Model call for a stable particle with a mass in the range of 100 GeV -- 1 TeV, most probably the lightest neutralino ($\mathrm{\chi_{1}^{0}}$), which is a natural cold dark matter candidate particle.
Since neutralino self-annihilation dominantly proceeds into \emph{continuum} gamma rays from jet-fragmentation processes, a low energy-threshold is crucial for disentangling the relatively soft spectra of astrophysical sources from the hard annihilation-debris spectrum. A high point source sensitivity also allows to distinguish time variable emission (e.g. due to AGN activity in the observed object) from the steady emission due to neutralino annihilations. The expected flux due to neutralino annihilation from a dark matter halo is
\begin{equation}
\mathrm{\Phi_{\gamma}(E)=\frac{1}{4\pi}\times\frac{\frac{1}{2}\left\langle\sigma v\right\rangle }{m_{\chi}^{2}}\times\int\rho^{2}_{\chi}\times\kappa\left[E,z\right] \times df\,[E(1+z)]\;\frac{c\,dt}{dz}\;dz}\label{eq:1}\;\;,\end{equation}
where $\mathrm{df\,[E(1+z)]}$ is the differential energy distribution of
the produced \linebreak photons per annihilation event, 
$\mathrm{\left\langle \sigma v\right\rangle }$
is the total self-annihilation cross section averaged over the thermal distribution of
the dark matter particles, and $\mathrm{m_{\chi}}$ is the particle mass. $\mathrm{\kappa\left[E,z\right]}$
parameterizes the gamma ray attenuation on cosmological scales \cite{kneiske}, and $\mathrm{\int}$$\rho^{2}_{\chi}\,\mathrm{ds}$ with $\mathrm{ds=\frac{c\,dt}{dz}\;dz}$
is the integral along the line of sight over the dark matter mass density
squared. Due to the strong foreground from starlight and cosmic ray interactions with interstellar hydrogen, observations at high galactic latitudes should have higher sensitivities than observations in the galactic plane. Since the energy threshold of an Imaging Air shower Cherenkov Telescope depends on the zenith angle, the Galactic Center is further disfavored as a target for telescopes on the northern hemisphere, unless extremely massive neutralino candidates are considered \cite{whipplegc}. On the other hand, the density-squared dependence of the annihilation induced flux qualifies nearby dwarf-spheroidal galaxies (e. g. Draco and Sagittarius) and the center of the Virgo Cluster as promising targets for indirect SUSY searches \cite{baltz}.

The most severe background for IACT observations naturally arises from hadronic air showers. Thus, any detection of the relatively weak expected dark matter signal will have to rely on the significance accumulated over a prolonged observation time and effective gamma-hadron separation.
Dedicated deep-field studies using MAGIC, e.g. of the central region of the Virgo Cluster, should therefore provide an unprecedented sensitivity for the detection of gamma rays due to neutralino dark matter annihilation. We use the dark matter halo profile of M 87 determined from X-ray observations of the hot intracluster gas in the center of the Virgo Cluster \cite{tsai}, and compare the expected sensitivity of MAGIC (at an energy threshold of 30 GeV) in an 800-hour pointing to M 87 with the flux predictions from a scan of the MSSM parameter space using the DarkSusy numerical routines and current accelerator limits \cite{darksusy}. 

We conclude that - aided by the flux enhancement due to substructure on mass scales down to $\mathrm{10^{8}}$ solar masses - MAGIC will indeed be able to probe a substantial part of the MSSM parameter space (Fig. 1). Note that the halo profile we employ is limited by the resolution of the X-ray observations, and is thus quite shallow in the inner part of the halo. If the halo profile of M 87 is indeed of the cusped type that is favored by high-resolution simulations of structure formation (see e. g. \cite{structure,st2}), the flux from M 87 will be significantly higher than predicted here, thus enlarging the accessible part of the MSSM parameter space. The same is true for the case of a significant enhancement of the dark matter densities at the centers of massive halos owing to adiabatic compression, which has also been recently discussed \cite{flix}.
Future extensions of MAGIC, i. e. ECO-1000, will substantially enhance the sensitivity and lower the energy threshold, further increasing the detection probability.
\begin{figure}[tb]
	\begin{center}
		\includegraphics{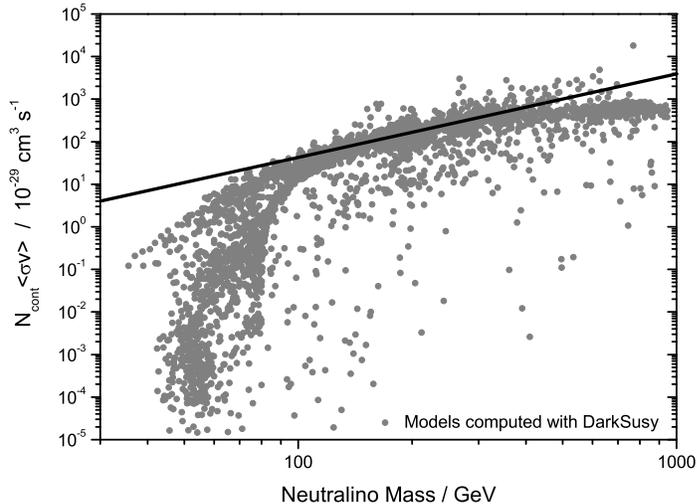}
	\end{center}
	\label{fig:ucla1}
	\caption{Scatter plot representing a scan over the MSSM parameter space using the DarkSusy numerical routines. The models are required to thermally produce $\mathrm{0.025\leq \Omega_{\chi}h^{2}\leq 0.176}$. For models that do not produce $\mathrm{\Omega_{\chi}h^{2}=\Omega_{DM}h^{2}}$ with $\mathrm{\Omega_{DM}h^{2}=0.116}$ and $\mathrm{h=0.71}$ \cite{wmap} we rescale $\mathrm{N_{cont}\left \langle \sigma v \right \rangle}$ with a factor of $\mathrm{\Omega_{\chi}h^{2}/0.116}$. The solid line denotes the $\mathrm{3 \sigma}$-limit for moderate amounts of substructure (lower mass cutoff for the subhalos: $\mathrm{10^{8}}$ solar masses), an 800-hour pointing of MAGIC and a threshold of 30 GeV.}
\end{figure}
\section{Extragalactic Gamma Ray and Neutrino Backgrounds due to Neutralino Annihilations}
If the nonbaryonic dark matter - neutralino assignment is correct, for indirect\\ supersymmetry-search strategies using gamma ray telescopes to work, the annihilation signal from the observed target must have a large contrast to the cosmological background signal from unresolved density fluctuations, which also emit gamma rays due to neutralino annihilations. One could argue that knowledge of the measured extragalactic gamma ray background alone would be sufficient. Deeper observations however will eventually reveal the coarse-grained structure of this background, which is different whether its origin lies in faint AGN or cosmologically distributed dark matter halos. To calculate this cosmological background of annihilation products, we start with the finite number of neutralinos produced during thermal freeze out in the early Universe, which are subsequently depleted following the Boltzman Equation $\;\mathrm{dn_{\chi}/dt=-\left\langle \sigma v\right\rangle\;n_{\chi}^{2}\,(1+z)^{3}}$ \cite{elsaesser,beu}. Taking into account the structure formation history \cite{ts} and cosmological gamma ray attenuation \cite{kneiske}, these calculations show that - owing to the strong $\mathrm{\rho^2}$-dependence of the signal - individual targets like M 87 will indeed stand out very clearly from this irreducible background. The background itself, and also the corresponding cosmological neutrino background, constitutes a prediction which may turn out to be interesting for studies of indirect dark matter signatures. Due to the strong clustering of the signal, this gamma ray background should trace the anisotropic mass distribution in the local universe, a signature which could be searched for using the survey-capability of the GLAST satellite. As can be concluded from Fig. 2, cuspy halo profiles and a high $\mathrm{\langle \sigma v\rangle}$ of $\mathrm{\mathcal{O}(10^{-25}\,cm^{3}\,s^{-1})}$ in combination with a neutralino rest mass of $\mathrm{m_\chi=(650-950)\,GeV}$ could well lead to a dominant high-energy component in the extragalactic gamma ray background at energies of a few GeV, which is indeed suggested by a new evaluation of the EGRET measurements \cite{strong}.
 
The corresponding extragalactic neutrino background is weak compared to the atmospheric background in the (1--100) GeV regime, but it can be comparable to other extragalactic neutrino backgrounds due to AGN sources and cosmic ray interactions in external galaxies. To suppress the overwhelmingly strong atmospheric background and single out the dark matter signal, a flavor-discriminating $\mathrm{\tau}$-neutrino detector would be required. Discrimination of $\mathrm{\tau}$-neutrinos in the GeV--TeV - regime might be achieved by looking for a "double-bang"-signature, albeit in the time structure of the PMT response rather than in the distribution of light along the track.
\begin{figure}[tb]
	\begin{center}
		\includegraphics{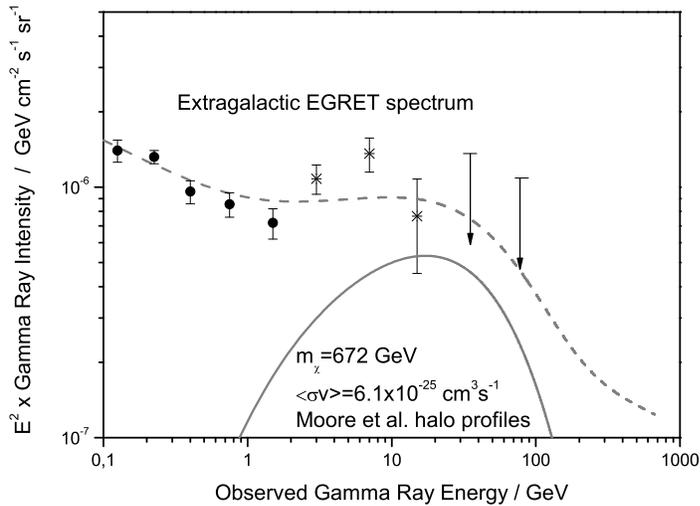}
	\end{center}
	\label{fig:ucla2}
	\caption{Extragalactic gamma ray background due to cosmological neutralino annihilation. The solid line represents the cosmological intensity from dark matter annihilations, while the dashed line shows this signal summed to a steep ($\mathrm{\alpha=-2.3}$) power law component presumably due to faint and unresolved Active Galactic Nuclei. The dots / stars with error bars and the upper limits represent the measured EGRET background from \cite{strong} and \cite{sreekumar}.}
\end{figure}
\section{Discussion}
In summary, we demonstrate that the search for gamma ray signatures of dark matter annihilation has entered an interesting era in which low energy-threshold telescopes like MAGIC or ECO-1000 could play an important role. Discriminating the possible SUSY-induced hard continuum flux from the astrophysical background remains the main challenge. Therefore, for dedicated search efforts, exact characterisation of the target and potential backgrounds is essential. These topics are addressed by the MAGIC Dark Matter Working Group.
Cosmological neutralino annihilation for both gamma rays and neutrinos leads to a potentially significant extragalactic continuum background, whose signatures render it an interesting topic of further study. The SUSY-induced spectral feature could be imprinted on the measured extragalactic gamma ray background.
\begin{ack}
D. E. and K. M. acknowledge support by the BMBF grant O5CM0MG1. Special thanks for help in preparing this contribution go to the entire MAGIC Dark Matter Working Group, especially H. Bartko, A. Biland, J. Flix, F. Pauss, S. Stark, R. Wagner and W. Wittek.
\end{ack}


\begin{thebibliography}{00}
\bibitem{magic_overview}M. Martinez et al., Proc. of the 28th ICRC, Tsukuba (2003)
\bibitem{bretz}T. Bretz, D. Dorner and R. Wagner, Proc. of the 28th ICRC, Tsukuba (2003)
\bibitem{first_alphas}T. Schweizer et al., "Status of the MAGIC Telescope", DPG Frühjahrstagung Teilchenphysik 2004, Mainz (2004)
\bibitem{eco}C. Baixeras et al., astro-ph/0403180 (2004)
\bibitem{eth}A. Biland, D. Elsässer, J. Flix, B. Moore, F. Pauss et al., ETH Zürich CDM Workshop (2004)
\bibitem{kneiske}T. M. Kneiske et al., Astron. and Astrophys., \textbf{413}, 807--815 (2004)
\bibitem{whipplegc}K. Kosack et al. astro-ph/0403422 (2004)
\bibitem{baltz}E. A. Baltz et al., Phys. Rev. D \textbf{61}, 2 (2000)
\bibitem{tsai}J. C. Tsai, Astrophys. J. \textbf{413}, L59 (1993)
\bibitem{darksusy}L. Bergström, J. Edsjö, P. Gondolo, P. Ullio, E. A. Baltz and M. Schelke,
$\mathrm{http://www.physto.se/\sim edsjo/darksusy/}$
\bibitem{structure}J. F. Navarro, C. S. Frenk and S. D. M. White, Astrophys. J. \textbf{462},
563 (1996)
\bibitem{st2}B. Moore, F. Govemato, T. Quinn, J. Stadel and G. Lake, Astrophys.
J. \textbf{499}, L5 (1998)
\bibitem{flix}F. Prada et al., astro-ph/0401512 (2004)
\bibitem{wmap}C. L. Bennett et al., ApJS \textbf{148}, 1 (2003)
\bibitem{elsaesser}D. Elsässer and K. Mannheim, submitted to Astropart. Phys. (2004)
\bibitem{beu}L. Bergström, J. Edsjö and P. Ullio, Phys. Rev. Lett. \textbf{87},
25 (2001)
\bibitem{ts}J. Taylor and J. Silk, MNRAS \textbf{339}, 505--514 (2003)
\bibitem{strong}A. W. Strong, I. V. Moskalenko and O. Reimer, Proc. of the 28th ICRC, Tsukuba (2003), and private communication
\bibitem{sreekumar}P. Sreekumar et al., Astrophys. J. \textbf{494}, 523--534 (1998)
\end{thebibliography}
\end{document}